\documentclass[%
 rsi,
 amsmath,amssymb,
 reprint,%
 groupedaddress,%
 superscriptaddress,
]{revtex4-1}

\usepackage{graphicx}
\usepackage{dcolumn}
\usepackage{bm}
\usepackage{booktabs}

\usepackage{xcolor} 
\usepackage[caption=false,font=footnotesize]{subfig}

\usepackage[utf8]{inputenc}
\usepackage[T1]{fontenc}
\usepackage{mathptmx}

\newcommand{\vecr}{\mathbf{r}}
\newcommand{\vecrd}{\mathbf{r_d}}

\newcommand{\K}{\mathbf{K}}

\newcommand{\Kn}{\mathbf{K_N}}
\newcommand{\A}{\mathbf{A}}

\newcommand{\An}{\mathbf{A_N}}
\newcommand{\M}{\mathbf{M}}
\newcommand{\Mn}{\mathbf{M_N}}

\begin{document}

\preprint{AIP/123-QED}

\title[]{On the robustness of model-based algorithms for photoacoustic tomography: \\
comparison between time and frequency domains}

%

\author{L. Hirsch}
\affiliation{Universidad de Buenos Aires, Facultad de Ingenier\'ia, Paseo Col\'on 850, C1063ACV, Buenos Aires, Argentina.}%
\author{M. G. Gonz\'alez}%
\email{The author to whom correspondence may be addressed: mggonza@fi.uba.ar}%
\affiliation{Universidad de Buenos Aires, Facultad de Ingenier\'ia, Paseo Col\'on 850, C1063ACV, Buenos Aires, Argentina.}%
\affiliation{Consejo Nacional de Investigaciones Cient\'ificas y T\'ecnicas, (CONICET), Godoy Cruz 2290, C1425FQB, Buenos Aires, Argentina.}%
\author{L. Rey Vega}%
\affiliation{Universidad de Buenos Aires, Facultad de Ingenier\'ia, Paseo Col\'on 850, C1063ACV, Buenos Aires, Argentina.}%
\affiliation{Consejo Nacional de Investigaciones Cient\'ificas y T\'ecnicas, (CONICET), Godoy Cruz 2290, C1425FQB, Buenos Aires, Argentina.}%

\date{\today}

\begin{abstract}
For photoacoustic image reconstruction, certain parameters such as sensor positions and speed of sound  have a major impact in the reconstruction process and must be carefully determined before data acquisition. Uncertainties in these parameters can lead to errors produced by a modeling mismatch, hindering the reconstruction process and severely affecting the resulting image quality. Therefore, in this work we study how modeling errors arising from uncertainty in sensor locations affect the images obtained by matrix model-based reconstruction algorithms based on time domain and frequency domain models of the photoacoustic problem. The effects on the reconstruction performance with respect to the uncertainty in the knowledge of the sensors location is compared and analyzed both in a qualitative and quantitative fashion for both time and frequency models. Ultimately, our study shows that the frequency domain approach is more sensitive to this kind of modeling errors. These conclusions are supported by numerical experiments and a theoretical sensitivity analysis of the mathematical operator for the direct problem. \\

\textit{This work has been submitted to the Review Scientific Instruments for publication.}
\end{abstract}

\maketitle

\section{Introduction}
\label{sect:intro}  

Photoacoustic tomography (PAT) is a non-invasive hybrid imaging modality based on the photoacosutic (PA) effect. By using  laser excitation and acoustic detection, PAT takes advantage of the high contrast imaging present in purely optical modalities while maintaining the great resolution given by ultrasonic detection \cite{xu2006}. The PAT technique relies upon the generation of ultrasonic acoustic waves (or PA waves) induced by the illumination of biological tissue with short-pulsed non-ionizing laser light. The absorbed energy leads to a rapid increase in temperature and to the formation of pressure waves due to a transient thermoelastic expansion of illuminated object. The PA waves travel through the tissue and are sensed by a series of ultrasonic transducers placed at a number of fixed positions around the sample. The received signals can then be analyzed and processed to recover the initial pressure distribution. Optical absorption is closely related to several important physiological properties, such as oxygen saturation and hemoglobin concentration. Thus, in past few decades, PAT has been successful in obtaining high fidelity images of vascular anatomy in small animals and functional images of blood oxygenation \cite{zhang2009, lutzwieler2013}. 
Unfortunately, the PAT inverse problem (the reconstruction of the 2-D or 3-D initial pressure distribution from the PA signals) is typically ill-posed. This means that even small errors in the measurements or inaccuracies in the modeling of the system can lead to an erroneous solution or a significantly distorted image \cite{tian2020}. These modeling errors have different sources and can occur due to several reasons: non-homogeneous speed of sound, the presence of noise in the measurements, sensor responses not sufficiently broadband or not omni-directional, limited-view detection, undersampling in space or time due to hardware constraints, etc \cite{hauptmann2020}. Furthermore, algorithms require precise information of the geometry of the experimental setup used for detecting the PA waves. Therefore, inaccuracies in the knowledge of the position, shape and size of the ultrasonic transducers can also lead to faulty modeling \cite{,tian2020,Sahlstrom2020}. 

Previously, some studies have analyzed how acoustic heterogeneities \cite{tick2019, matthews2018, dean2012, jin2006} and finite transducer size \cite{mitsuhashig2014,rosenthal2011} affect the solution of the inverse problem. The effect of different discretizations of the forward problem on image reconstruction was briefly discussed in \cite{rosenthal2010}. More recently, Salhström et al. \cite{Sahlstrom2020} studied the usefulness of using a Bayesian framework for taking into account some of the inherent modeling errors in PAT when approaching the inverse problem.

Several reconstruction algorithms have been employed in the estimation of the initial pressure distribution \cite{lutzwieler2013, Sahlstrom2020}. A distinction between analytic inversion formulas and algebraic inversion procedures, both in time- and frequency-domain, can be made \cite{rosenthal2013}. Analytic reconstruction techniques use the mathematical model of the PA pressure propagation in order to perform the exact inversion of mathematical operator of the forward problem \cite{Minghua_Xu_Wang_2002, and_Wang_2002a, and_Wang_2002b}. Algebraic reconstruction techniques differ conceptually from analytic formulas as the inversion is typically performed from a numerical approximation to the forward or inverse problem. The most relevant approaches of each group are the Universal Back-Projection \cite{xu2005} and model-based algorithms \cite{guo2010}, respectively. 
The back-projection (BP) algorithms are based on approximate analytical inversion formulas, and they are closely related with the (inverse) spherical Radon transform \cite{tick2016}. Due to this, BP algorithms are limited to a specific reconstruction geometry (e.g. cylindrical \cite{burgholzer2007}) and require that the PA signals be densely sampled along the detection surface \cite{rosenthal2013}. Furthermore, practical implementations of these closed-form solutions may lead to the appearance of significant streak-type artifacts, blurring and negative values in the reconstructed images \cite{xu2005, haltmeier2010}. 
In contrast, the matrix model-based (MM) image reconstruction algorithms are based on a discrete representation of the acoustic forward model describing the propagation of pressure and not on a particular analytical solution like BP type approaches. This allows the construction of a matrix, which jointly represents the forward operator and the PAT measurement system and which is strongly dependent on the characteristics of the experimental setup. With this approach, the image reconstruction is performed by numerically minimizing the error (usually quantified by a quadratic loss function) between the measured acoustic signals and those predicted using the acoustic forward model \cite{rosenthal2013}. Typically, the minimization is done through a least squares based inversion of the model matrix. As this matrix could be ill-conditioned, regularization techniques such as Tikhonov regularization are typically used to stabilize the numerical inversion \cite{tick2016}. MM frameworks tend to be more computationally intensive than BP algorithms, since they require calculating and manipulating large matrices \cite{ding2016}. Nevertheless, they can be more versatile given that they can be applied to arbitrary measurement geometries and that many additional linear effects can be added to the model \cite{rosenthal2010}.

The accuracy or fidelity of the model used for representing an experimental setup is crucial for the success of MM algorithms \cite{rosenthal2011}. Since the construction of the matrix depends heavily on certain parameters of a given measurement geometry, care must be taken in the characterization of the PAT system to ensure good image reconstruction quality \cite{tian2020}. However, in typical measurement setups, eliminating all the sources that give rise modeling errors can be a difficult task \cite{Sahlstrom2020}. For example, temperature variations during the measurement process may cause slight drifts in the speed of sound of a medium \cite{tick2019} or the measurement setup may be built in such a way that the exact determination of transducers locations could result challenging \cite{Sahlstrom2020}.
Hence, it is of interest to analyze the behaviour of MM algorithms when some of the measurement setup parameters suffer from uncertainties in their values. In particular, such an analysis may show if some MM schemes prove to be more robust than others, or if some lead to better overall image quality when faced with uncertain parameters.

In this paper we analyze how uncertain knowledge of sensor locations in a typical experimental setup can lead to unsatisfactory image reconstruction quality due to modeling errors. Two common MM algorithms are tested: one working in the time domain \cite{paltauf2002} and another in the frequency domain \cite{provost2009}. Through simulations, uncertainty is added to the model matrices in a controlled manner, allowing us to analyze qualitatively and quantitatively the quality of the resulting reconstructed images. In particular, it is shown that it is advantageous to use a time domain over a frequency domain model when the knowledge of the sensor positions becomes moderately uncertain. In other words, the time domain model proves to be more robust to modeling errors than the frequency domain model when the value of a key parameter in the measurement setup like the transducer locations, is not accurately known. 
This same behaviour is then observed when attempting to reconstruct an image from experimental measurements from a PAT setup in which the sensor positions are only known to $\sim1\%$ of their nominal value. Moreover, we discuss how uncertanties in the speed of sound in an acoustically homogeneous medium leads to a similar behaviour for the frequency domain MM method.

The remainder of this paper is organized as follows. Sections \ref{subsec:met_forward} and \ref{subsec:met_inverse} introduce the forward and inverse acoustic problems in PAT. Sections \ref{subsec:metsim} and \ref{subsec:metexp} explain the methodology for the simulations and experimental reconstructions. Then, in section \ref{s:results}, the obtained results are presented. In section \ref{s:disc}, a theoretical sensitivity analysis of the mathematical operator for the direct problem is provided. Finally, the conclusions of this paper are in section \ref{s:conclu}. 

\section{Methods}
\label{s:met}

\subsection{Forward problem}
\label{subsec:met_forward}

According to PA theory, for an acoustically non-absorbing homogeneous medium the acoustic pressure $p(\mathbf{r},t)$ at position $\mathbf{r} \in\mathbb{R}^3$ and time $t$ originated from the optical absorption of a sample excited by an electromagnetic pulse $\delta(t)$, satisfies the homogeneous wave equation \cite{burgholzer2007}:

\begin{equation}
\left(\frac{\partial^2}{\partial t^2} - v_s^2 \, \nabla^2 \right) p(\mathbf{r},t) = 0
\label{eq:waveeq}
\end{equation}

\noindent with the initial conditions,

\begin{equation*}
p(\mathbf{r},0) = p_0(\mathbf{r})\, \text{,} \quad \left(\partial p /\partial t\right)(\mathbf{r},0)= 0 
\label{eq:waveeq_cond_ini}
\end{equation*}

\noindent where $p_0(\mathbf{r})$ is the initial OA pressure and $v_s$ represents the speed of sound in the medium. Moreover, if thermal and acoustic confinement are fulfilled, that is, when the laser pulse duration is short enough that the heat conduction and acoustic propagation into neighboring regions of the illuminated region can be neglected, the initially induced pressure $p_0(\mathbf{r})$ is proportional to the total absorbed optical energy density \cite{lutzwieler2013}. 

By solving the wave equation using the Green’s function in free space, the forward solution of the PA pressure detected by an ideal point-detector at position $\mathbf{r_d}$ placed on a surface $S$ surrounding the volume of interest $V$, can be written as \cite{biomedicalOptics}:     

\begin{equation}
p_d(\mathbf{r_d},t)=\frac{1}{4\pi\,v_s^2} \frac{\partial}{\partial t}\iiint_{V} \, p_0(\mathbf{r}) \frac{\delta\left(t-|\mathbf{r_d}-\mathbf{r}|/v_s\right)}{|\mathbf{r_d}-\mathbf{r}|} d^3\mathbf{r}
\label{eq:fo_time}
\end{equation}

\medskip

The goal of the PAT inverse problem is to reconstruct the initial pressure $p_0(\mathbf{r})$ from the measured PA signals $p_d(\mathbf{r_d},t)$ at different positions $\mathbf{r_d}$ on a surface $S$ surrounding the volume of interest \cite{lutzwieler2013}.

Taking the Fourier transform with respect to $v_s\,t$, the forward operator (\ref{eq:fo_time}) can be rewritten in the frequency domain as \cite{xu2006}: 

\begin{equation}
\bar{p}_d(\mathbf{r_d},k) = -i\, k \iiint_V \, p_0(\mathbf{r}) \, \frac{\exp\left(-i\,k\, |\mathbf{r_d}-\mathbf{r}|\right)}{4\pi|\mathbf{r_d}-\mathbf{r}|}\, d^3\mathbf{r}
\label{eq:fo_frec}
\end{equation}

\noindent where $k=\omega/v_s$ , $\omega$ is the angular frequency equal to $2\pi f$, and $f$ is the signal temporal frequency. 

\subsection{Acoustic inverse problem}
\label{subsec:met_inverse}

Analytical inversion formulas were used for image reconstruction tasks in most PA systems in the past \cite{dean2013,rosenthal2013}. The reconstruction approach most favored in several articles are BP type algorithms, due to their simple implementation and applicability to most practical imaging scenarios \cite{rosenthal2013}. As mentioned in section \ref{sect:intro}, one of the most prominent formulations of the BP approach is the universal back-projection algorithm \cite{xu2005}. In a homogeneous medium with a constant $v_s$, the universal back-projection formula directly links $p_0(\mathbf{r})$ to $p_d(\mathbf{r_d},t)$ on the detection surface $S$ that encloses the PA source \cite{xu2005}:

\begin{equation}
p_0(\mathbf{r})=\int_{\Omega_s} b\left(\mathbf{r_d},t=|\mathbf{r_d}-\mathbf{r}|/v_s\right) \, \frac{d\Omega_s}{\Omega_s}
\label{eq:ubp}
\end{equation}
\medskip

\noindent where $b(\mathbf{r_d},t)=2\,p_d(\mathbf{r_d},t)-2\,t\,\partial p_d(\mathbf{r_d},t)/\partial t$ is the back-projection term related to the measurement at position $\mathbf{r_d}$, $\Omega_s$ is the solid angle of the whole surface $S$ with respect to the reconstruction point inside $S$, $d\Omega_s = dS \,\cos \theta_s/|\mathbf{r_d}-\mathbf{r}|$ and $\theta_s$ denotes the angle between the outwards pointing unit normal of $S$ and $(\mathbf{r_d}-\mathbf{r})$. The above formula is exact for cylindrical, planar and spherical geometries, and assumes point detectors with infinite bandwidth and uniform angular response \cite{cui2021,rosenthal2013,rosenthal2010}. However, in practice, the transducers can not be considered point-like, are band limited and do not have flat angular responses. In these non-ideal imaging scenarios, the analytical inverse formulas significantly deviate from reality, generating imaging artifacts (e.g. appearance of negative values that have no physical interpretation) and distorted images \cite{rosenthal2013,rosenthal2010}. 

A different reconstruction approach is based on the MM algorithms. In this technique, the forward solution is represented by a matrix equation, which is used for solving the inverse problem. One of the advantages of this approach is that any linear effect in the system may be considered. Therefore, any spatio-temporal detection response that can be modeled or measured, may be taken into account in the inversion process  \cite{paltauf2018,rosenthal2013,paltauf2002}. In the ideal case of point detectors and a homogeneous lossless acoustic medium, the model matrix in the time domain may be calculated by discretizing the integral relation in (\ref{eq:fo_time}) \cite{rosenthal2013}:

\begin{equation}
\mathbf{p_d}=\mathbf{A} \, \mathbf{p_0}
\label{eq:mbt}
\end{equation}

\medskip

\noindent where $\mathbf{p_d} \in\mathbb{R}^{N_d \cdot N_t\times 1}$ is a column vector representing the measured pressures at a set of detector locations $\mathbf{r_d}_l$ ($l=1 \ldots N_d$) and time instants $t_k$ ($k=1 \ldots N_t$); $\mathbf{p_0} \in\mathbb{R}^{N\times 1}$ is a column vector representing the values of the initial acoustic pressure on the imaging region grid; and $\mathbf{A} \in\mathbb{R}^{N_d \cdot N_t\times N}$ is the model matrix. The $j$-th element ($j=1 \ldots N$) in $\mathbf{p_0}$ contains the average value of the initial pressure within a volume element of size $\Delta V$ at position $\mathbf{r}_j$. Once the discrete formulation has been established, the inverse problem is reduced to the algebraic problem of inverting (\ref{eq:mbt}). The matrix $\mathbf{A}$ can be written as the multiplication of two matrices $\mathbf{A^{oa} \, A^s}$ where $\mathbf{A^s}$ represents the point response function of the imaging system for an ideal point-like sensor and $\mathbf{A^{oa}}$ is a time derivative operator. The matrix $\mathbf{A^s}$ is defined as \cite{paltauf2018, wang2011,dean2012,guo2010}: 

\begin{equation}
    A^s_{lkj} = \frac{1}{4\pi v_s^2}\frac{\Delta V}{\Delta t^2} \frac{d(t_k,\mathbf{r}_j)}{|\mathbf{r_d}_l - \mathbf{r}_j|}
\label{eq:Gs1}
\end{equation}

\begin{equation}
    d(t_k,\mathbf{r}_j) = \begin{cases}
    1 & \text{si } |t_k - \frac{|\mathbf{r_d}_l - \mathbf{r}_j|}{v_s}| < \Delta t/2  \\
    0 & \text{else } 
  \end{cases}
    \label{eq:Gs2}
\end{equation}

\medskip

\noindent where $\Delta t$ is the time step at which $p_d(\mathbf{r_d},t)$ are sampled and $\mathbf{A^{oa}}$ is a square matrix with replications of the basic ``N-shape'' signals grouped in columns along its main diagonal \cite{paltauf2018}.

In the frequency domain, the matrix representation of the forward model is similar as the time domain model described above \cite{provost2009,meng2012_2}. From (\ref{eq:fo_frec}) we can obtain:

\begin{equation}
\mathbf{\bar{p}_d}=\mathbf{K} \, \mathbf{p_0}
\label{eq:mbf}
\end{equation}

\medskip

\noindent where $\mathbf{\bar{p}_d} \in\mathbb{C}^{N_d \cdot N_f \times 1}$ is a column vector representing the measured pressures at $\mathbf{r_d}_l$ positions ($l=1 \ldots N_d$) and frequency samples $f_p$ ($p=1 \ldots N_f$); and $\mathbf{K} \in\mathbb{C}^{N_d \cdot N_f\times N}$ is the model matrix in the frequency domain which can be written as \cite{provost2009,meng2012_2}:

\begin{equation}
K_{lpj} = -i\,k_p \, \Delta V\, \frac{\exp\left(ik_p\,|\mathbf{r_d}_l - \mathbf{r}_j|\right)}{4\pi|\mathbf{r_d}_l - \mathbf{r}_j|}
\label{eq:Km}
\end{equation}

\noindent where $k_p = 2\,\pi \,f_p/v_s$.

In the case of a ﬁnite-size detector, the spatial impulse response (SIR) of the sensor must be taken into account \cite{rosenthal2011}. For the numerical calculation of the SIR, the area of the detector is divided into surface elements (treated as point detectors) which are then added up. Secondly, in \eqref{eq:Gs1} and \eqref{eq:Km}, a weight factor to take into account the size of surface elements and the directional sensitivity of the detectors is included \cite{paltauf2018}.

As already mentioned above, once a model matrix for the ideal case is constructed, the MM approach allows the model matrix to be modified in order to refine the forward model, such as to include the frequency or time response of the detector or the time/frequency properties of the illumination pulse \cite{paltauf2018,rosenthal2013}. Furthermore, the MM approach enables imposing constraints on the PA source to regularize the solution of the inverse problem. A typical regularization scheme is the so called Tikhonov regularization, which involves a square error minimization criterion coupled with a term weighing the $\ell_2$ norm of the solution \cite{provost2009,dean2012}:

\begin{equation}
\mathbf{\hat{p}_0} = \min_{p_0} \: || \mathbf{A} \, \mathbf{p_0} - \mathbf{p_d} ||_{\ell_2}^2 + \alpha \, ||\mathbf{p_0}||_{\ell2}^2 
\label{eq:regTik}
\end{equation}

\medskip

\noindent where $\alpha\geq 0$ is the regularization parameter. Given a fixed $\alpha$, the solution to \eqref{eq:regTik} is unique and is called the Tikhonov-regularized pseudo-solution \cite{provost2009}:

\begin{equation}
\mathbf{\hat{p}_0} = (\mathbf{A}^H\,\mathbf{A} + \alpha \mathbf{I})^{-1} \mathbf{A}^H\,\mathbf{p_d}
\label{eq:solTik}
\end{equation}

\medskip
\noindent where $\mathbf{I}$ is the identity matrix and $H$ denotes the conjugate transpose operator. The value of the regularization parameter $\alpha$ can affect the solution and must be carefully chosen. For example, higher regularization tends to over-smooth the image while lower $\alpha$ values amplifies the noise in the images. In order to obtain the inverse equations for the frequency domain, $\mathbf{A}$ and $\mathbf{p_d}$ should be replaced by $\mathbf{K}$ and $\mathbf{\bar{p}_d}$  on (\ref{eq:regTik}) and (\ref{eq:solTik}).

\subsection{Simulations}
\label{subsec:metsim}
\begin{figure}
\centering
\includegraphics[width = 0.35\textwidth]{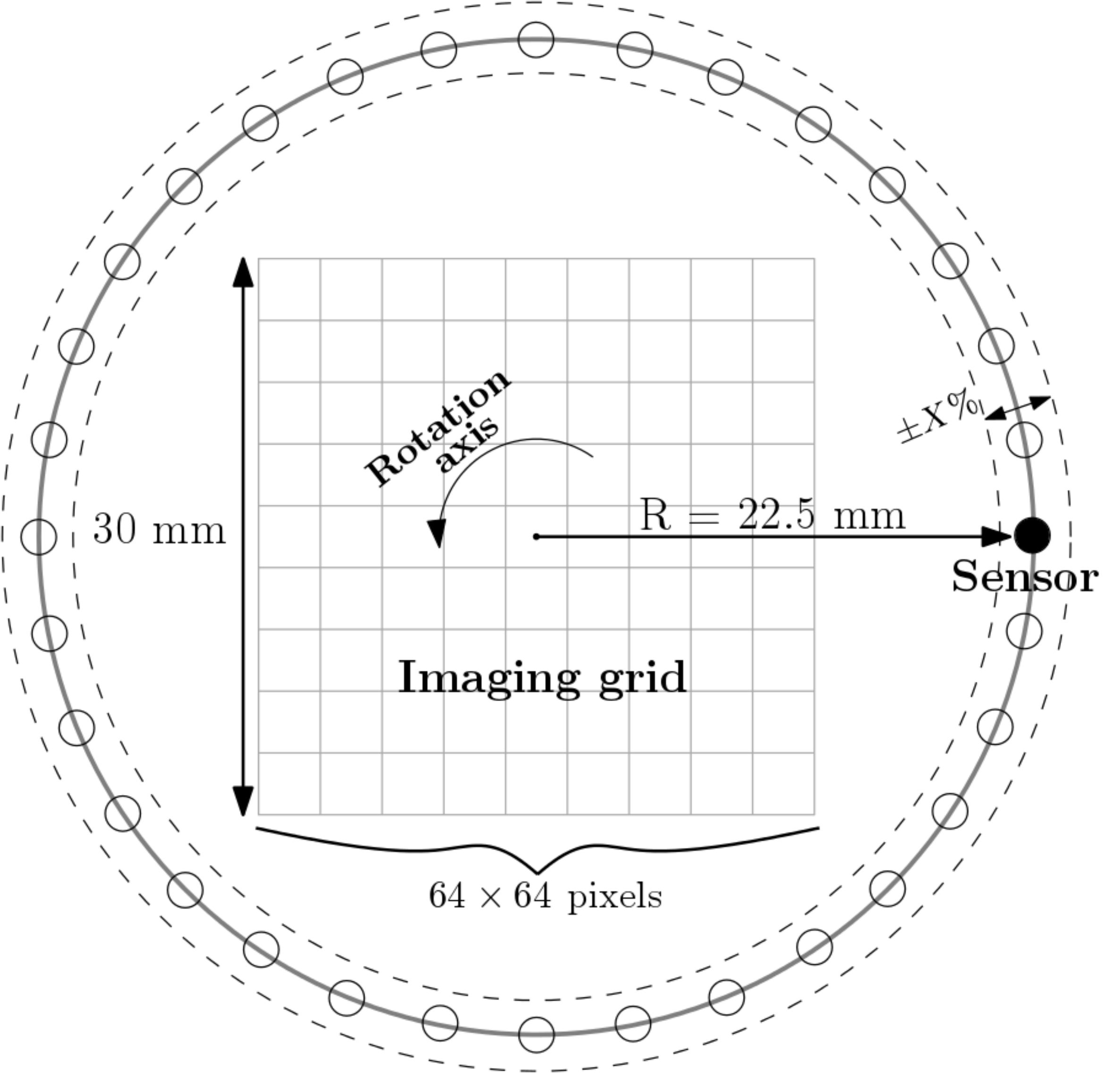}
\caption{Simulation geometry of the PA data acquisition setup.}
\label{fig:simulsetup}
\end{figure}

The effect of uncertainties in ultrasound sensor locations was studied using the simulation geometry of the PA data acquisition setup shown in Fig. \ref{fig:simulsetup} with increasing levels of uncertainty. The setup presented in Fig. \ref{fig:simulsetup} represents a two-dimensional PAT system implemented with a sensor rotating around the imaging region where the sample, uniformly illuminated, is placed. This single-detector based PAT system was demonstrated to be very useful for proof-of-concept studies due to its simplicity, low cost, and effectiveness \cite{tian2020,sharma2019}. A square imaging region having a size of 30 mm x 30 mm and a resolution of 64 x 64 pixels was used and the sensor (black circle) was placed on a circle of 22.5 mm radius. The PA signals were detected over 120 locations placed equidistantly around this circumference. Although large area detectors are typically used in practice, for simplicity, we assumed the sensor to be point-like with a bandwidth of 20 MHz. For data collection, the time step $\Delta t$ was 50 ns with $N_t=N_f=$ 420 samples. The speed of sound was set to $v_s=$ 1500 m/s and the medium was assumed homogeneous with no absorption or dispersion of sound. The transducer frequency response was modeled using a band-pass filter with upper and lower cutoff frequencies 0.1 MHz and 20 MHz, respectively. In order to simulate data with uncertainties in sensor locations, the ultrasound sensor locations $\mathbf{r_d}$ were perturbed as illustrated in Fig. \ref{fig:simulsetup}. These altered sensor locations were drawn from generalized uniform distributions for the radial values. The simulation conducted can be summarized into three steps: 

\begin{enumerate}
    \item Construct $\A$ and $\K$ with a radial value $R=$ 22.5 mm. Generate $\mathbf{p_d}$ and $\mathbf{\bar{p}_d}$ from a phantom $\mathbf{p_0}$ using these model matrices. 
    \item Construct $\An$ and $\Kn$ using a radial value in the uniform interval $\left[R - X\%,\, R + X\%\right]$, where $X$ represents the uncertainty introduced in the radius of the circumference of the sensor location. $\An$ and $\Kn$ are the matrices assumed by the reconstruction algorithm to be the real ones.
    \item Solve (\ref{eq:regTik}) employing the data obtained in step 1 and the model matrices constructed in step 2 with an iterative method to approximate the solution, using the \textit{lsqr} algorithm provided in the Python module \textit{scipy.sparse.linalg}.
\end{enumerate}

In order to perform a statistical analysis, for each value of $X$, we performed 50 reconstructions with different radial sensor location randomly obtained within the range $\left[R - X\%,\, R + X\%\right]$. 

As the initial pressure distribution $\mathbf{p_0}$, we used the standard Shepp–Logan numerical phantom \cite{shepp1974}. In all cases, Gaussian uncorrelated noise with zero mean and standard deviation of 1$\%$ of the maximum simulated peak amplitude, resulting in a signal to noise ratio (SNR) of 40 dB, was added to the data. Moreover, for comparison, we also reconstructed the image using the BP approach based on \eqref{eq:ubp} \cite{xu2005}. The simulations were carried out in the Python framework. 

\subsection{Experimental setup}
\label{subsec:metexp}
\begin{figure}
\centering
\includegraphics[width = 0.46\textwidth]{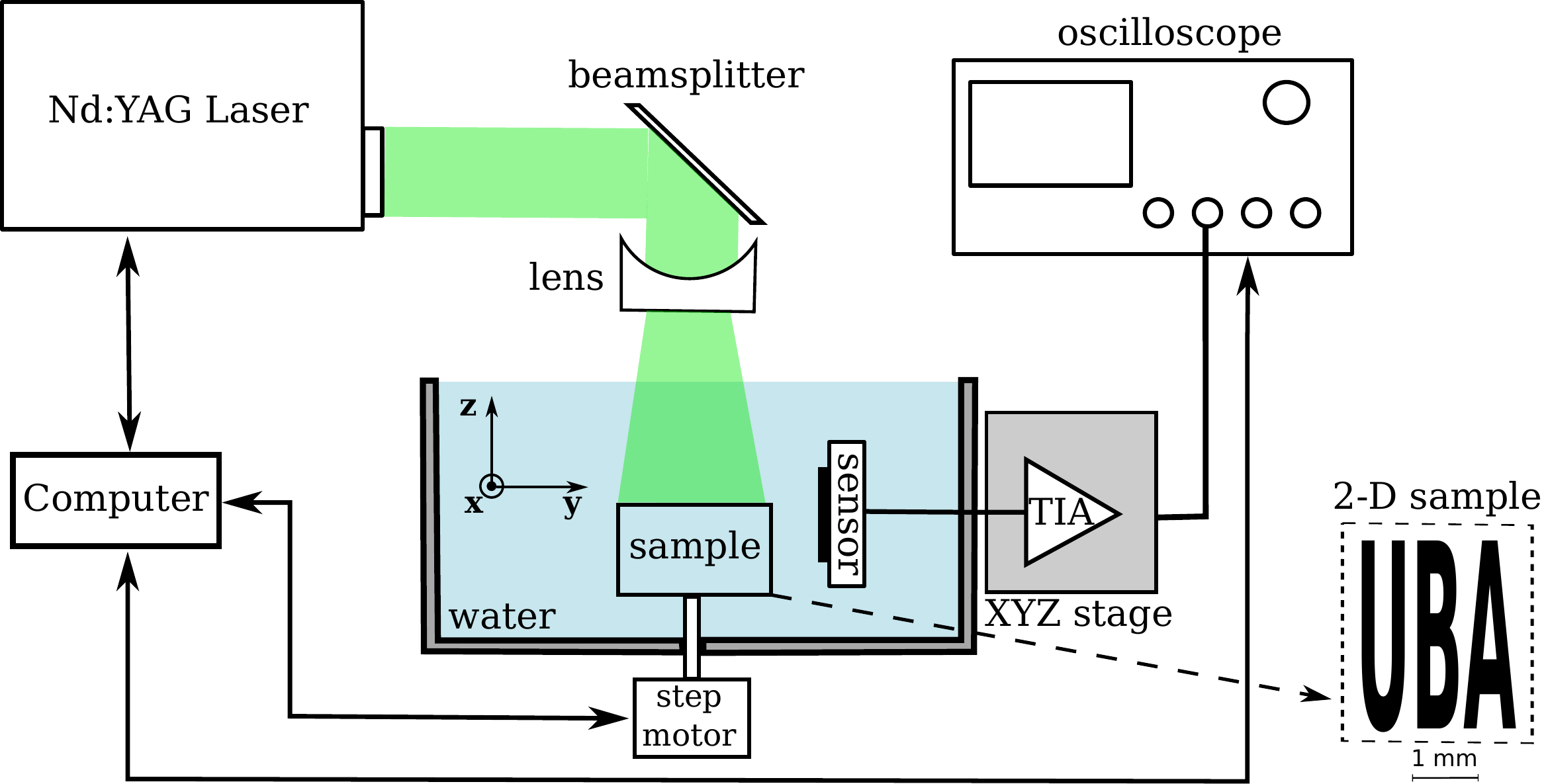}
\caption{Experimental setup.}
\label{fig:expsetup}
\end{figure}

In order to obtain experimental PA signals for comparing the performance between the different reconstruction algorithms under study, we used a 2-D PAT system whose capabilities have been demonstrated in a previous work \cite{insabella2020b}. A diagram of the experimental setup is presented in Fig. \ref{fig:expsetup}. A polymer piezoelectric sensor and a sample were immersed in a vessel filled with deionized water. The water temperature was measured with a calibrated thermocouple to determine the speed of sound ($v_s=1479$ m/s). A Nd:YAG laser with second harmonic generation (Continuum Minilite I, 532 nm), 5 ns pulse duration, 10 Hz repetition rate and pulse energy less than 1 mJ, was the light source. A diverging lens adapted the diameter of the laser beam to a size larger than the sample, trying to achieve homogeneous illumination. The ultrasonic detector was fixed and pointed to the center of the rotating sample stage using an $xyz$ translation stage and it was oriented along the $z$ axis, perpendicular to the image plane $xy$. The distance between the sensor and the center of the rotating sample was (8.35 $\pm$ 0.05) mm (a position uncertainty of about 1\%). The phantom was fixed to a rotatory stage (Newport PR50CC) and rotated 360$^{\circ}$ in 3$^\circ$ steps. The sensor output was amplified with a transimpedance amplifier (FEMTO HCA-100MHz-50K-C), digitized by an oscilloscope (Tektronix TDS 2024) and processed on a personal computer. The PA signals were recorded and averaged 16 times. A pyroelectric detector (Coherent J-10MB-LE) measured the laser pulse energy.

The ultrasonic sensor used in this experiment is an integrating line detector with an active detection area of approximately 0.7 mm x 24 mm. The detection system (sensor + amplifier) has a pressure sensitivity of  1.6 mV/Pa, a noise equivalent pressure (NEP) of 0.4 mPa/$\sqrt{\text{Hz}}$ and a bandwidth of 20 MHz \cite{insabella2020b}.

The sample consists of an ink pattern laser (three capital letters reading UBA) printed on a transparent film embedded in agarose gel.  The agarose gel was prepared with 2.5$\%$ (w/v) agarose in distilled water. First, a cylindrical base of the agarose gel with a diameter of 14 mm and a height of $\sim$20 mm was prepared. Then, the object (ink pattern on transparent film) was placed in the middle of the cylinder and fixed with a few drops of the gel. Finally, another layer of gel with a thickness of $\sim$1 mm was formed on top of the sample object. A picture of the sample is shown in Fig. \ref{fig:experiment}.a. 

\section{Results}
\label{s:results}

\subsection{Simulations}
\label{subsec:simul}

\begin{figure}
\centering
\includegraphics[width = 0.46\textwidth]{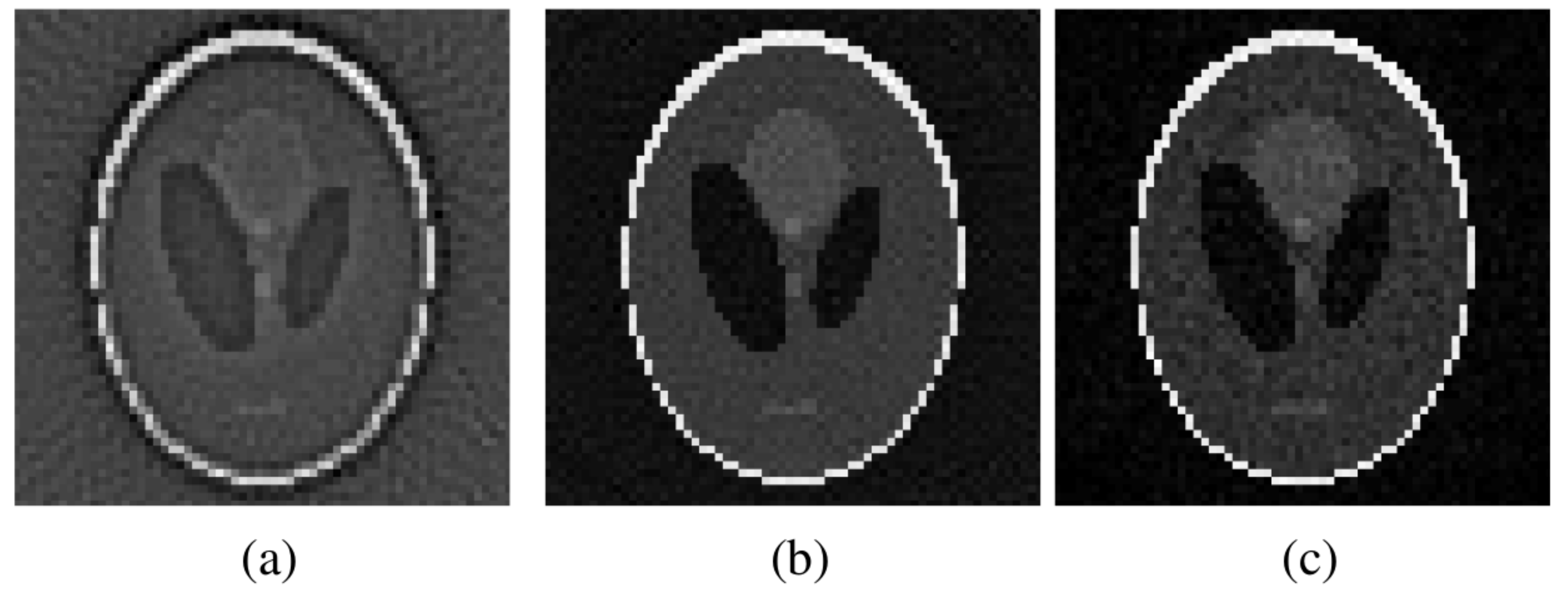}
\caption{Average images obtained for $X=0.1\%$. (a) BP. (b) TDMM. (c) FDMM.}
\label{fig:sim_01percent}
\end{figure}

\begin{figure}
\centering
\includegraphics[width = 0.46\textwidth]{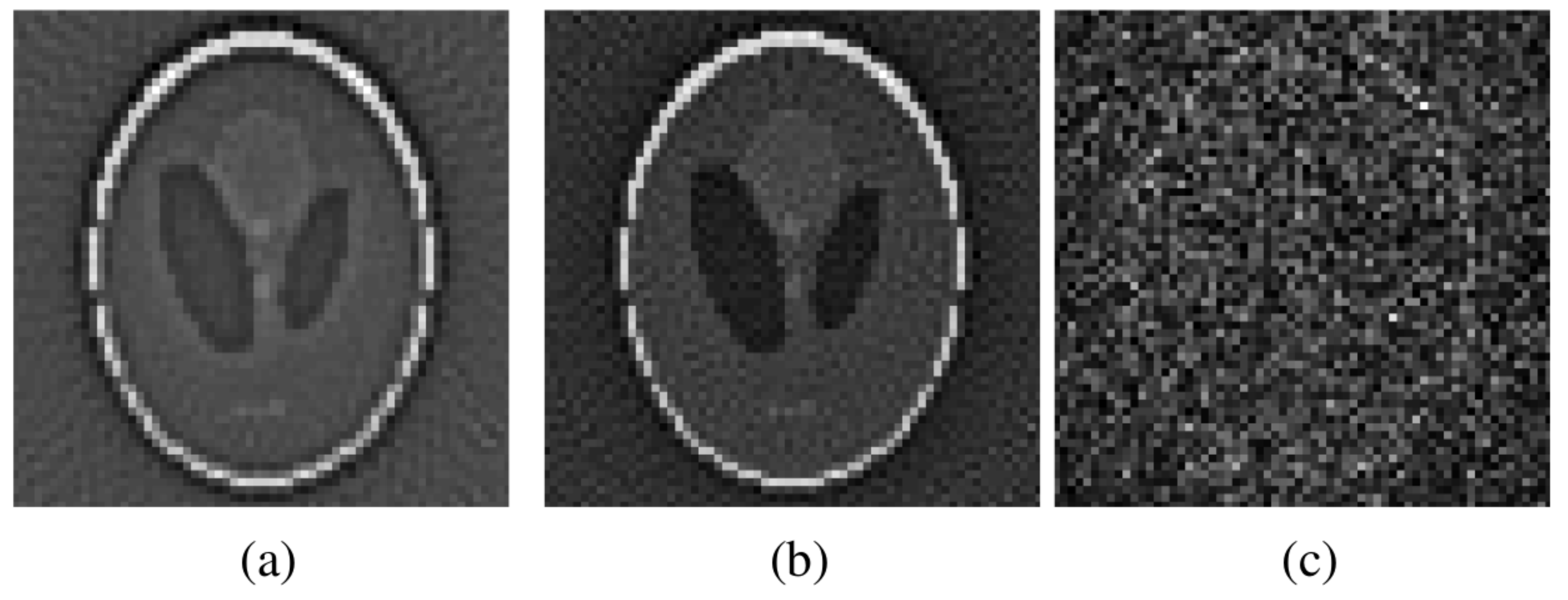}
\caption{Average images obtained for $X=1\%$. (a) BP. (b) TDMM. (c) FDMM. }
\label{fig:sim_1percent}
\end{figure}

\begin{figure}
\centering
\includegraphics[width = 0.46\textwidth]{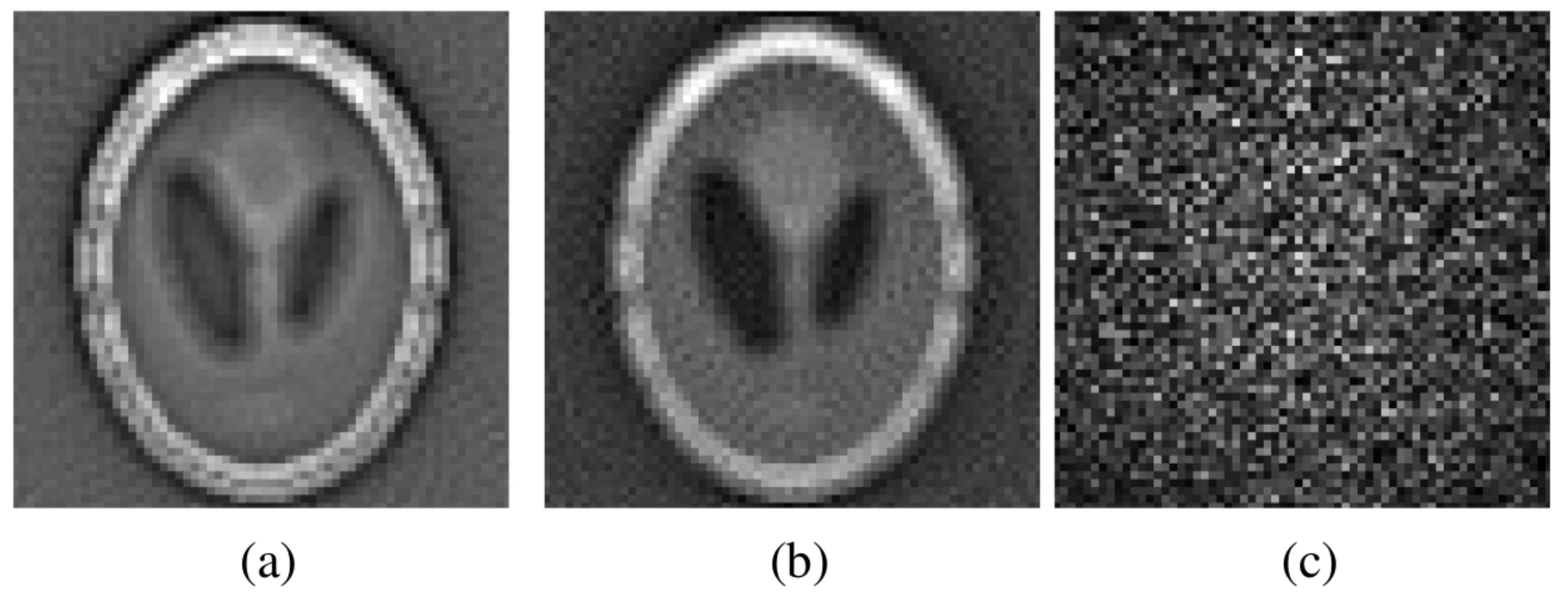}
\caption{Average images obtained for $X=5\%$. (a) BP. (b) TDMM. (c) FDMM.}
\label{fig:sim_5percent}
\end{figure}

Following the steps detailed in section \ref{subsec:metsim}, we simulated radial uncertainties $X\%$ in the range between 0.01 $\%$ and 10 $\%$. Figs. \ref{fig:sim_01percent}, \ref{fig:sim_1percent} and  \ref{fig:sim_5percent} show the averaged images obtained by the MM approaches for three different radial uncertainties values in the range under study. As a benchmark, the reconstructed images using the BP method based on (\ref{eq:ubp}) are also presented.


\begin{figure}
\centering
\includegraphics[width = 0.45\textwidth]{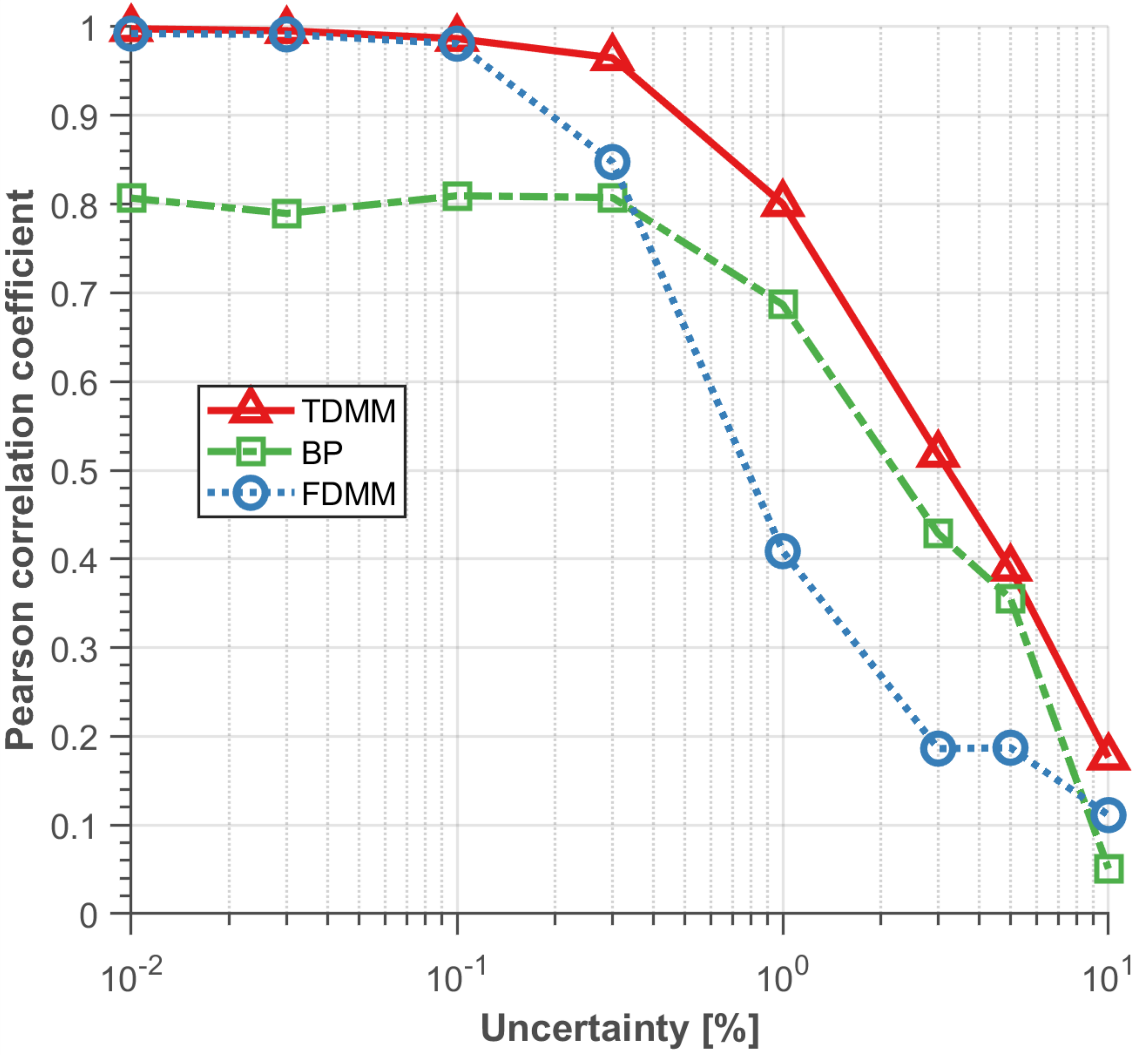}
\caption{Correlation value between the reconstructed image and the original image as a function of the uncertainty $X$.}
\label{fig: PSgraph}
\end{figure}

\begin{figure*}
\centering
\includegraphics[height=4.9cm]{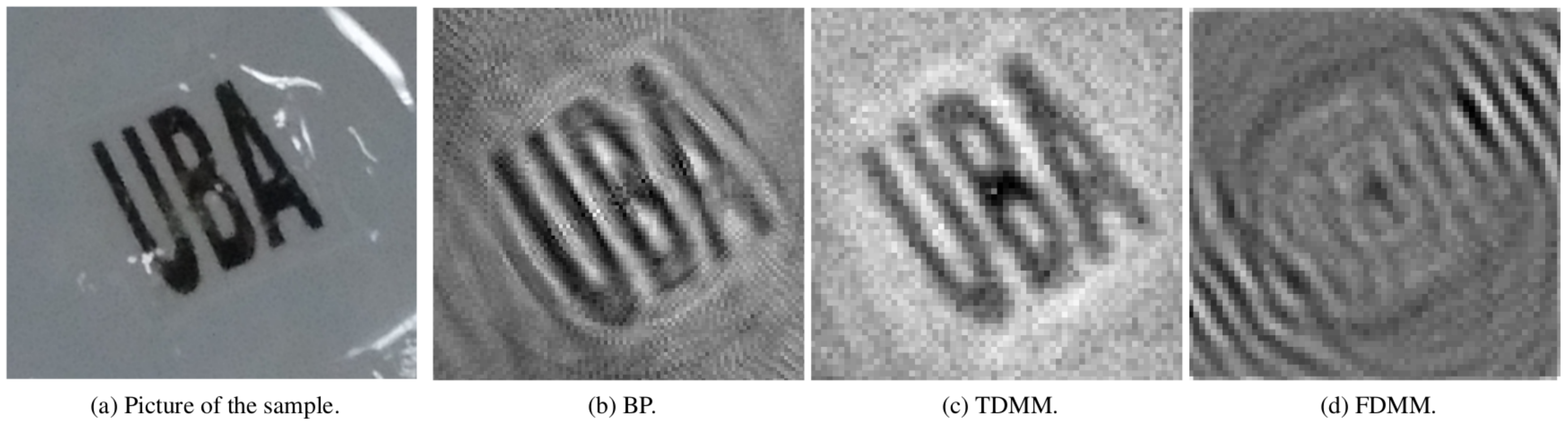}
\caption{Images obtained employing the different reconstruction algorithms using 120 locations with $\sim1\%$ sensor position uncertainty. The phantom (ink pattern in agarose gel) reads UBA and its size is 3.0 mm x 2.5 mm. The calculated PC values are: (b) PC$=0.612$; (c) PC$=0.723$; and (d) PC$=0.123$.}
\label{fig:experiment}
\end{figure*}

As it can be seen in Fig. \ref{fig:sim_01percent}, the phantom can be perfectly recognized in the reconstructed images for $X= 0.1 \%$. However, as the value of $X$ increases, the images obtained with the frequency domain model matrix (FDMM) method are more affected than reconstructions obtained by the time domain model matrix (TDMM) and BP approaches. For example, in the image obtained by the frequency approach for $X= 1 \%$ (Fig. \ref{fig:sim_1percent}.c) almost no appreciable detail of the original image can be appreciated. On the other hand, the images reconstructed using time domain methods (Figs. \ref{fig:sim_1percent}.a and \ref{fig:sim_1percent}.b) are comparable to that achieved for $X= 0.1 \%$.  For higher values of $X$, the images obtained with the time domain approaches are also affected as shown on Fig. \ref{fig:sim_5percent}. However, even for $X= 5 \%$, we are still able to observe a diffuse version of the phantom reconstructed with TDMM and BP (Figs. \ref{fig:sim_5percent}.a and \ref{fig:sim_5percent}.b), while, for the FDMM approach, it is not possible to observe any vestige of the original image (Fig. \ref{fig:sim_5percent}.c). 

In order to carry out a quantitative performance analysis of the reconstructed images, we use the Pearson Correlation (PC) coefficient as an evaluation metric. The PC is a measure of linear correlation between two entities, here, the original image and the reconstructed image \cite{prakash2014}. It has a value between -1 and 1, with either extreme signifying perfect correlation (or anti-correlation) and a value of 0 meaning no correlation between the images. 
Fig. \ref{fig: PSgraph} shows the calculated PC values versus the simulated radial uncertainties $X\%$ (on a logarithmic scale) for the three reconstruction methods. It can be observed that the value of the PC remains relatively constant up to an uncertainty of 0.3 \% for the time domain methods (TDMM and BP) and begins to fall sharply after $X\%$ exceeds 1\%. In the case of the FDMM algorithm, the PC is approximately constant up to 0.03 \% and starts to decrease rapidly for $X\%$ values greater than 0.1 \%. 

\subsection{Experiment}
\label{subsec:experiment}
In order to verify the results of the simulations, we carried out measurements with the 2-D PAT system detailed in section \ref{subsec:metexp}.  As the sensor of the experimental setup is not a point detector, we made modifications to the reconstruction algorithms used in the simulations. In the MM approach, for the numerical calculation of the SIR of the detector, its area is divided into surface elements which are then added up. Moreover, in order to take into account the size of surface elements and the directional sensitivity of the detectors, a weight factor is included in (\ref{eq:Gs1}) and (\ref{eq:Km}) \cite{paltauf2018}. 

On the other hand, based on \cite{burgholzer2007}, we implemented a temporal BP algorithm for integrating line detectors. The reconstructed images from the measured pressures with a radial location uncertainty of $\sim$1 \%  are presented in Fig. \ref{fig:experiment}. As it is mentioned in \ref{subsec:experiment}, the location uncertainty arises from $xyz$ translation stage used to position the sensor.
The reconstructions presented on Fig. \ref{fig:experiment} show a behavior similar to that found in section \ref{subsec:simul} where the time domain methods obtain a better quality image than FDMM. With TDMM (Fig. \ref{fig:experiment}.c) and BP (Fig. \ref{fig:experiment}.b) it is possible to distinguish the capital letters UBA in the center of the image. However, with FDMM, it is not possible to see any detail of the original phantom (Fig. \ref{fig:experiment}.d). The failure of the latter method can be attributed to the uncertainty in the sensor position. The time domain approaches also show some artifacts characteristic of the methods themselves. For example the BP (Fig. \ref{fig:experiment}.b) show some streak type artifacts caused by the finite number of angles used in the reconstruction \cite{rosenthal2013}. On the other hand, the fuzzy edges seen in the image corresponding to TDMM (Fig. \ref{fig:experiment}.c) are due to the effect of the Tikhonov regularization \cite{provost2009}. The calculated PC values of this images, indicated in the caption of Fig. \ref{fig:experiment}, are in agreement with the values presented in Fig. \ref{fig: PSgraph}. 

\section{Discussion}
\label{s:disc}
As observed via simulations in section \ref{subsec:simul}, the FDMM approach showed higher sensitivity to slight variations in sensor positions than time domain methods. We attributed this sensitivity to the failure of the FDMM approach in recovering an image using experimentally obtained data in section \ref{subsec:experiment}. This results may indicate that when using a traditional pulsed laser experimental setup as the one used in section \ref{subsec:metexp}, time domain reconstruction techniques may be preferred over frequency domain model based approaches.

A possible explanation accounting for the greater sensitivity of the FDMM algorithm comes from looking at the entries of the matrix $\K$ given by \eqref{eq:Km}. One of the terms in \eqref{eq:Km} is a complex exponential whose argument is $i\,\frac{f_p\,2\,\pi}{v_s}\,|\mathbf{r_d} - \mathbf{r}|$. In a typical PAT setting, the frequency values $f_p$ range from a few hundred kHz to several tens of MHz. Therefore, any small error in the value of the sensor position $\mathbf{r_d}$ is amplified through its multiplication with $f_p$, leading to a drastic change in the value of the high frequency complex exponential (or phasor). This phenomenon can be shown through a simple numerical example. Taking $f_p = 1\,\text{MHz}$, $v_s = 1500\,\text{m/s}$ and $\vecr =( 0\, \hat{x} + 0\, \hat{y} + 0\, \hat{z})\,\text{mm}$, if a detector is placed on a point with coordinates $\mathbf{r_d} = 22.500 \, \hat{x} \, \text{mm}$, the complex exponential evaluates to $1.00 + i\,0.00$. Now suppose that, due to a measurement error, the detector position is considered to be at $\widetilde{\mathbf{r_d}} = 22.275 \, \hat{x}\,\text{mm}$ (only a 1\% difference with respect to $\mathbf{r_d}$). Using $\widetilde{\mathbf{r_d}}$, the value of the complex exponential becomes $0.59 - i\,0.81$. 
Finally, the term $\frac{\exp\left(i\,f_p\,2\,\pi\,|\mathbf{r_d} - \mathbf{r}|\right)}{|\mathbf{r_d} - \mathbf{r}|}$ in \eqref{eq:Km} computes to roughly $(44.44 + i\,0.00)\, \text{m}^{-1}$ when using $\mathbf{r_d}$ and to $(26.4-i\,36.3)\, \text{m}^{-1}$ when using $\widetilde{\mathbf{r_d}}$, a substantial difference. The frequency domain matrix $\K$ is highly dense (i.e not sparse) and every entry in it depends upon the value of the complex exponential present in \eqref{eq:Km}. As a consequence, any tiny error inside the high frequency phasor affects every value in $\K$, leading to a matrix that does not accurately represent the system at hand. This effect only gets worse with higher values of $f_p$, meaning that higher bandwidth systems will be much more sensitive to uncertainties in $\vecrd$. Nonetheless, even for typical detectors operating between  1-5 MHz, this simple numerical example shows that the error generated by this effect is noticeable.

On the other hand, the time domain model given by \eqref{eq:mbt}, whose characteristic matrix $\A$ is defined by \eqref{eq:Gs1}, is also affected by this problem but in a less extreme manner due to the lack of an exponential term. Instead, the term involving the detector position is $\frac{d(t_k,\mathbf{r})}{|\mathbf{r_d} - \mathbf{r}|}$ where $d(t_k,\mathbf{r})$ is a discrete representation of the delta term in \eqref{eq:fo_time} and can only be either 1 or 0, which does not drastically affect the numerical value of the matrix entry. Using the same numerical example as before, this term evaluates to $44.4\, \text{m}^{-1}$ when using $\mathbf{r_d}$ and $44.9 \, \text{m}^{-1}$ when using $\widetilde{\mathbf{r_d}}$, only a small difference. Slight changes in $\vecrd$ also correspond to a slight shift in the position of the matrix entries in $\A$ due to the $d(t_k,\mathbf{r})$ term. This tiny shifting of the matrix entries is the cause of the blurring observed in the TDMM reconstructions in Section \ref{subsec:simul}, since the pressures measured at the detectors are interpreted as originating from different (but nearby) pixel in the image.

Incidentally, note that since the speed of sound $v_s$ is present in the same complex exponential (for the FDMM algorithm) and delta term (for the TDMM one) as $\vecrd$, any uncertainties in this parameter will have a similar effect upon the matrices and lead to similar problems.

We can provide some more analysis regarding the sensitivity of frequency and time domain models. Both models are linear ones as equations (\ref{eq:mbt}) and (\ref{eq:mbf}) show. In the following we will denote with $\mathbf{M}$ the discretized forward operator present in those equations (either matrix $\mathbf{A}$ or $\mathbf{K}$). The presence of uncertainties in the measurement system implies the existence of a deviation in operator $\mathbf{M}$ with respect to a nominal situation. In more precise terms, the measured pressures at the detector locations can be written as:

\begin{eqnarray}
\mathbf{p}_\mathbf{d}&=&\M \,\mathbf{p}_{\mathbf{0}}+\mathbf{n}\nonumber\\
&=&\Mn \, \mathbf{p}_{\mathbf{0}}+\Delta\M \,\mathbf{p}_{\mathbf{0}}+\mathbf{n}
\label{eq:nominal_model}
\end{eqnarray}

\noindent where $\M$ is the actual discretized forward operator, $\Mn$ is the nominal model, assumed by the reconstruction algorithm to be the true one, $\Delta\M$ is the deviation from that nominal model induced by the uncertainties of the measurement system and $\mathbf{n}$ is a additive measurement noise (i.e., white noise in the detectors). From previous equation, it can be seen that the uncertainties of the measurement system can be interpreted as an additional source of noise, $\Delta\M \, \mathbf{p}_{\mathbf{0}}$, which is dependent of the ground truth $\mathbf{p}_\mathbf{0}$ in the forward model. This additional perturbation makes the reconstruction of the image harder. A simple characterization of the influence of this quantity is the norm of this term: $\|\Delta\M \, \mathbf{p}_{\mathbf{0}}\|$. Clearly, the precise influence of this quantity depends not only on $\Delta\mathbf{M}$, but also on the specific details of $\mathbf{p}_{\mathbf{0}}$. For example, if $\mathbf{p}_{\mathbf{0}}$, is aligned with a particular bad direction of $\Delta\M$, the influence of that term will be significant. From a practical point of view, it is better to have a quantitative measure of the influence of that term but also independent of the true value of $\mathbf{p}_\mathbf{0}$. In that sense, we can look for a worst-case criterion given by:

\begin{equation}
\|\Delta\mathbf{M}\|_2\equiv \sup_{\mathbf{p}_\mathbf{0}\neq\mathbf{0}}\frac{\|\Delta\mathbf{M}\mathbf{p}_{\mathbf{0}}\|}{\|\mathbf{p}_\mathbf{0}\|}
    \label{eq:spectral_norm}
\end{equation}

Equation (\ref{eq:spectral_norm}) is the mathematical definition of the spectral norm of matrix $\Delta\mathbf{M}$ and requires obtaining its singular value decomposition in order to be computed \cite{Golub_Loan_1996}. 

Finally, using these definitions, a quantitative measure of the effect of the uncertainties, with respect to the case of perfect knowledge of the measurement system (i.e., $\mathbf{M}=\mathbf{M}_N$) can be given by:

\begin{equation}
\delta \M \triangleq \frac{\|\Delta\M\|_2 }{\|\M\|_2 }=\frac{\|\M - \Mn\|_2 }{\|\M\|_2 }
    \label{eq:final_measure_uncertain_v2}
\end{equation}

The idea behind this metric is to allow us to understand what happens to model matrices $\A$ and $\K$ as the percentage error $X\%$ in the detectors position $\vecrd$ increases. The term $\|\M - \Mn\|_2$ in \eqref{eq:final_measure_uncertain_v2} quantifies the difference between the true forward operator $\M$ (the one that correctly models the system) and the operator $\Mn$ which contains the modeling errors (and it is assumed to be true by the reconstruction algorithm). Thus, if $\delta \M$ is low over increasing values of $X\%$, it would mean that $\Mn$ can accurately model the system even in the face of uncertainty in its transducers position. On the other hand if $\delta \M$ where to increase rapidly, it would mean that the model is sensitive to slight changes in $\vecrd$ since the difference term $\|\M - \Mn\|_2$ grows quickly. Intuitively, this means that using $\Mn$ to invert data would lead to a poor quality and highly distorted image for even relatively low values of $X\%$.

\begin{figure}
\centering
\includegraphics[width = 0.45\textwidth]{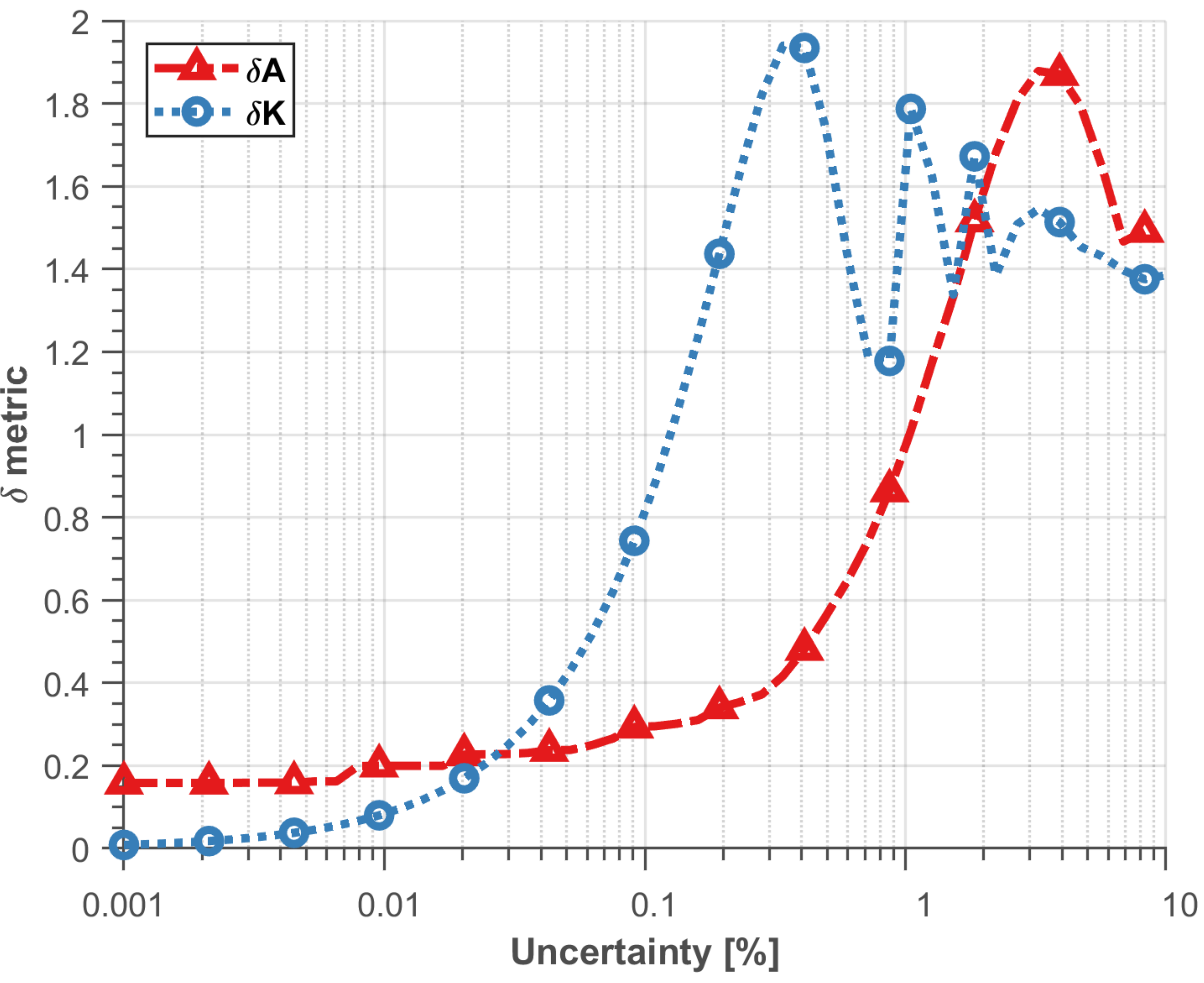}
\caption{$\delta$ metric (Eq. \eqref{eq:final_measure_uncertain_v2}) for TDMM and FDMM model matrices $\A$ and $\K$ as a function of the uncertainty in the transducer positions.}
\label{fig:Errormat}
\end{figure}

Computing the $\delta$ metric is straightforward through simulation. First, the true model $\A$ or $\K$ is constructed (via \eqref{eq:Gs1} or \eqref{eq:Km}) by using a fixed value for the detector position $\vecrd$. Then $\A_{\mathbf{N}}$ or $\K_{\mathbf{N}}$ is calculated with a corresponding detector position $\vecrd + X\%$, were $X\%$ is the percent error manually introduced to represent faulty modeling of the system. Finally, using the corresponding matrices, the $\delta$ metric is computed according to \eqref{eq:final_measure_uncertain_v2}.
Fig. \ref{fig:Errormat} shows a graph of $\delta \A$ and $\delta \K$ for increasing values of $X\%$ plotted on a logarithmic scale. 
As it can be clearly seen $\delta \K$ increases faster than $\delta \A$, indicating that the matrix entries present in $\K_\mathbf{N}$ are vastly different from those present in $\K$, even when the percent difference in the value in $\vecrd$ is small. This means that even a slight difference between the experimentally measured and true values of $\vecrd$ will lead to a matrix that can not accurately represent the system, and consequently, will not be able to reconstruct an good quality image. As mentioned before, we attribute this behaviour to the high frequency phasor present in \eqref{eq:Km}, whose value changes drastically given tiny changes in the value of $\vecrd$ used to construct the FDMM matrix. Fig. \ref{fig:Errormat} was obtained using the same bandwidth, $v_s$ and $\vecrd$ values as the simulations in Section \ref{subsec:simul}

Two other phenomena can be observed in Fig. \ref{fig:Errormat}. First, the variations in the value of $\delta \K$ after it reaches its peak value are due to the oscillatory nature of the complex exponential that make up the entries in $\K$. Secondly, both $\delta \A$ and $\delta \K$ will seem to decrease and converge to the same value when $X\%$ increases sufficiently. This can be easily explained recalling the definition for the $\delta$ metric \eqref{eq:final_measure_uncertain_v2} and the expressions for the matrix entries given in \eqref{eq:Gs1} and \eqref{eq:Km}. Both \eqref{eq:Gs1} and \eqref{eq:Km} contain a term of the form $1/|\vecrd + \vecr|$, which diminishes in value as $\vecrd$ increases, making every matrix entry smaller in absolute value. Thus, as $X\%$ increases, $\vecrd + X\%$ increases and the coefficients in $\Mn$ get smaller. As a result the value of $\|\M - \Mn\|_2$ will approach $\| \M \|_2$ making $\delta M$ approach 1 for large enough $X\%$.
It is interesting to note that, when uncertainties in the speed of sound of the medium are present, a qualitatively similar behaviour as in Fig. \ref{fig:Errormat} is obtained. Once again, as occurs with $\vecrd$, this is due to the presence of the speed of sound $v_s$ in the high frequency complex exponential in \eqref{eq:Km}. 

The authors acknowledge other articles in which the FDMM has been successfully applied. In \cite{provost2009, meng2012, meng2012_3} a setup with a pulsed laser is employed and image is obtained using a similar process as the one described in \ref{s:met}, but no detail is given on the relative uncertainties in the measurement systems. As shown in section \ref{subsec:simul}, if the uncertainty in the transducer position is low enough, little modeling error is introduced and both the time domain and frequency domain approaches are able to reconstruct good quality images. 
Other articles exist in which the FDMM has been applied successfully, but contain some noteworthy differences. For example, in \cite{mohajerani2014}, a frequency domain PAT system is presented in which the sample is excited by an amplitude modulated continuous-wave laser operating at different frequencies, allowing for the retrieval of phase and amplitude measurements.
This fundamentally changes the data processing and image reconstruction steps, and may be the key to a more robust setup. But again, no mention of the uncertainties in the system is given and a specialized setup different from the usual pulsed laser approach is used.

\section{Conclusions}
\label{s:conclu}

In this paper, the robustness of time and frequency domain model-based reconstruction algorithms was compared and analyzed. It was shown that in the presence of modeling errors caused by uncertain transducer positions in a typical PAT setup, the time domain approach proved to be more robust than the frequency domain one, which was unable to recover an image even when the uncertainty was moderately low.

The robustness of the different approaches was first studied through simulations, where modeling errors were deliberately added to the matrices used for reconstruction. This allowed for the qualitative and quantitative analysis of the obtained images and clearly showed the deterioration present in the frequency domain reconstructions. The experimental reconstructions done using a typical pulsed laser PAT setup with a $\sim1\%$ uncertainty in its transducers positions showed similar results.
The analysis provided in Section \ref{s:disc} supports the conclusion of the frequency domain approach being less robust and points to a high frequency complex exponential as a possible source for its sensitivity. The model mismatch caused by uncertainties can be interpreted as an additional source of noise, causing distorted reconstructions. All in all, the results indicate that when using a standard pulsed laser PAT setup, time domain reconstruction approaches should be used in favor of frequency domain ones, due to their robust characteristics. 

Even though the frequency domain approach suffers from severe robustness issues when model uncertainties are present, it has proven useful when employing more advanced image reconstruction algorithms, such as the application of the compressed-sensing paradigm PAT image reconstruction \cite{provost2009}. Such cases would require experimental setups with very low working uncertainty to obtain good quality images. This opens the path analysing more closely the frequency model's limitations and studying ways to improve its robustness, such as using a Bayesian approach \cite{Sahlstrom2020} for solving the inverse problem. Another possible alternative, more suited for the frequency domain approach, is to study the implementation of a signal processing based calibration approach in order to estimate the uncertainties in the measurement setup in a similar way to the procedures typically used in MIMO-OFDM communication systems \cite{Minn_2010}.

\subsection* {Acknowledgments}
This work was supported by the ANPCyT (grants PICT 2016-2204 and PICT 2018-04589) and by the UBA (UBACyT grants 20020190100032BA).





\bibliography{references}   


\end{document}